\documentclass{moriond}

\def\be{\begin{equation}}
\def\ee{\end{equation}}
\def\bea{\begin{eqnarray}}
\def\eea{\end{eqnarray}}

\newcommand{\Photo}{}

\begin{document}
\vspace*{4cm}
\title{Review of searches for new physics at CMS}

\author{Anne-Mazarine Lyon, on behalf of the CMS Collaboration} 

\address{ETH Z{\"u}rich, Institute for Particle Physics and Astrophysics\\
Otto-Stern-Weg 5, 8093 Z{\"u}rich, Switzerland}

\maketitle\abstracts{
A review of recent results from searches for new physics is presented. The analyses exploit data sets collected by the CMS experiment during Run~2 of the CERN LHC. Searches for exotic particles decaying into two bosons and for heavy neutral leptons, as well as a model-agnostic search based on anomaly detection are summarised. No significant sign for the presence of new physics was found. Prospects for the data parking strategy during Run~3 are also discussed. 
}

\section{Introduction}
The standard model (SM) of particle physics is a highly predictive and well-tested theoretical framework that describes the electroweak and strong interactions. However, it does not provide explanations for key questions, both observational and theoretical. Possible extensions of the SM that address those questions are commonly referred to as new physics. A large part of the physics program of the CMS experiment~\cite{CMS} at the CERN LHC is dedicated to searches for signs for new physics in the data. In this paper, an overview of recent results from these searches is presented. Diboson searches are discussed in Section~\ref{section_diboson}, while searches for heavy neutral leptons (HNLs), predicted as the heavy  partners of the SM neutrinos, are described in Section~\ref{section_HNL}. The first model-agnostic search at CMS based on anomaly detection is presented in Section~\ref{section_anomaly}. Finally, Section~\ref{section_parking} discusses the Run~3 data parking strategy and Section~\ref{section_summary} summarises the main results.

\section{Diboson searches}\label{section_diboson}

\begin{figure}[tbh!]
\centering
\includegraphics[width=0.45\textwidth]{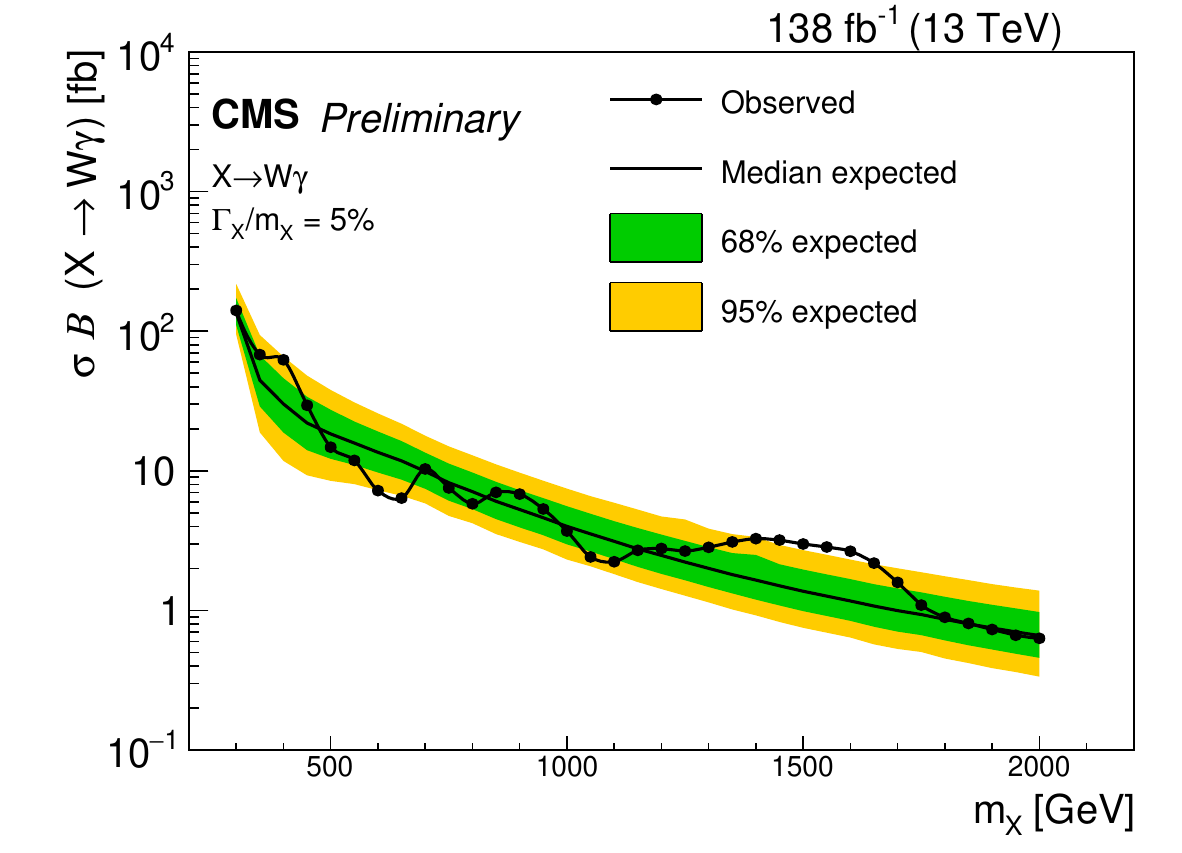}
\hspace{0.5cm}
\includegraphics[width=0.45\textwidth]{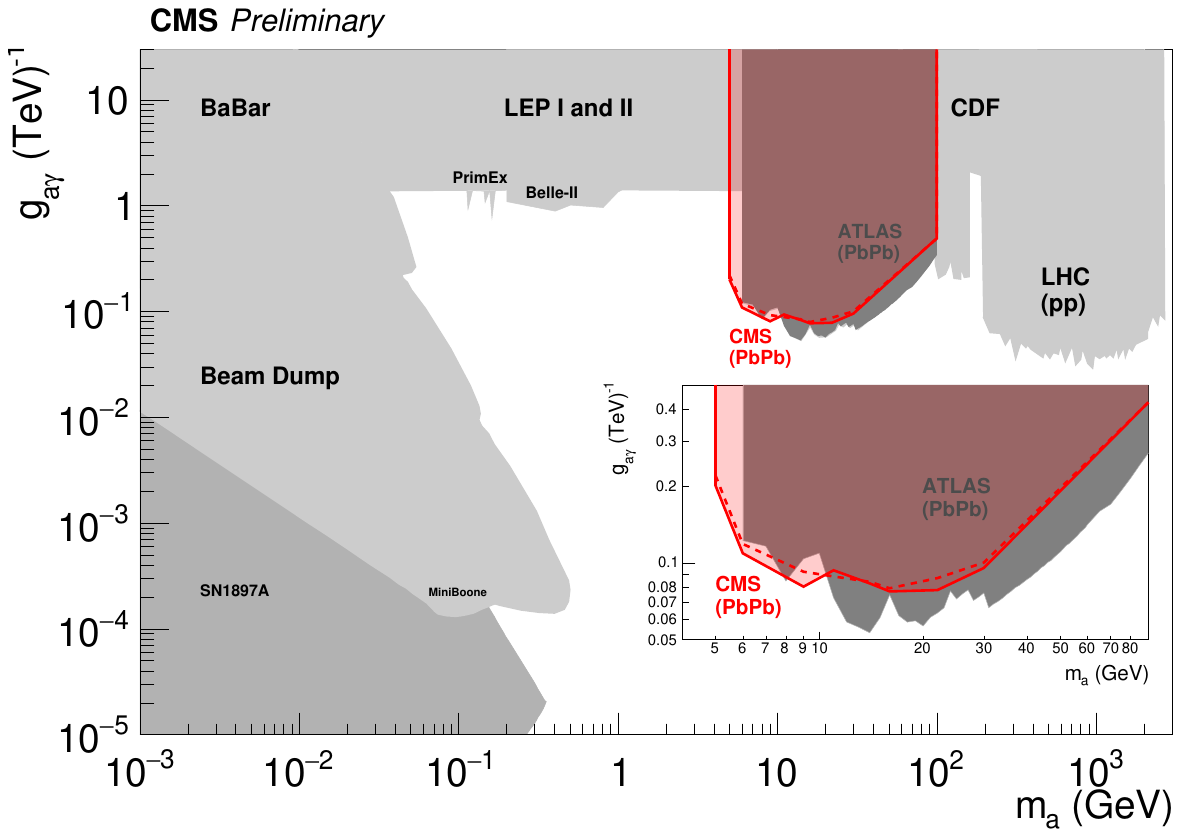}
\caption{\textit{(Left)} The 95\% CL limits on $\sigma\mathcal{B}(\mathrm{X}\to\mathrm{W}\gamma)$ obtained from the combination of the leptonic and hadronic analyses, under the broad resonance width assumption. \textit{(Right)} The 95\% CL limits on the axion-photon coupling, $g_{\mathrm{a}\gamma}$, shown as a function of the mass $m_\mathrm{a}$. The solid (dashed) red lines represent the observed (expected) limits. \label{figure_diboson}}
\end{figure}

The first analysis~\cite{EXO-21-017} is a search for a charged resonance, X, decaying into a W~boson and a photon, performed using the Run~2 proton-proton (pp) collision data ($\sqrt{s}=13$~TeV, 138~fb$^{-1}$). This analysis probes leptonic decays of the W boson into an electron or a muon, and is complementary to the early analysis targeting hardonic decays of the W boson~\cite{EXO-21-017_hadronic}. No significant deviation from the background-only prediction was observed. Figure~\ref{figure_diboson} (left) shows the upper exclusion limits at 95\% confidence level (CL) on the product of the X cross section and its branching fraction to W$\gamma$, as obtained from the combination of the leptonic and hadronic channels. These limits are the most stringent to date for the model under scrutiny. Moreover, the combination of the two analyses reduces from $3.1\sigma$ to $2.5\sigma$ the local significance of the deviation observed in the hadronic analysis at a mass $m_\mathrm{X}=1.58$~TeV.
  
The second analysis~\cite{HIN-21-015} seeks for axion-like particles, a, in ultraperipheral lead-lead collisions ($\sqrt{s_\mathrm{NN}}=5.02$~TeV, 1.65~nb$^{-1}$). The signal models assume that the exotic particle decays into two photons,  measured in the central part of the detector, and that the lead ions escape the detection at low angles with respect to the beam line. The observed data were found to be compatible with the SM background. The upper exclusion limits on the coupling to photons are the most stringent to date for masses $5<m_\mathrm{a} <10$~GeV, as can be seen in Fig.~\ref{figure_diboson} (right).

\section{Heavy neutral lepton searches}\label{section_HNL}

\subsection{Searches for HNLs in W boson decays}

\begin{figure}[b!]
\centering
\includegraphics[width=0.43\textwidth]{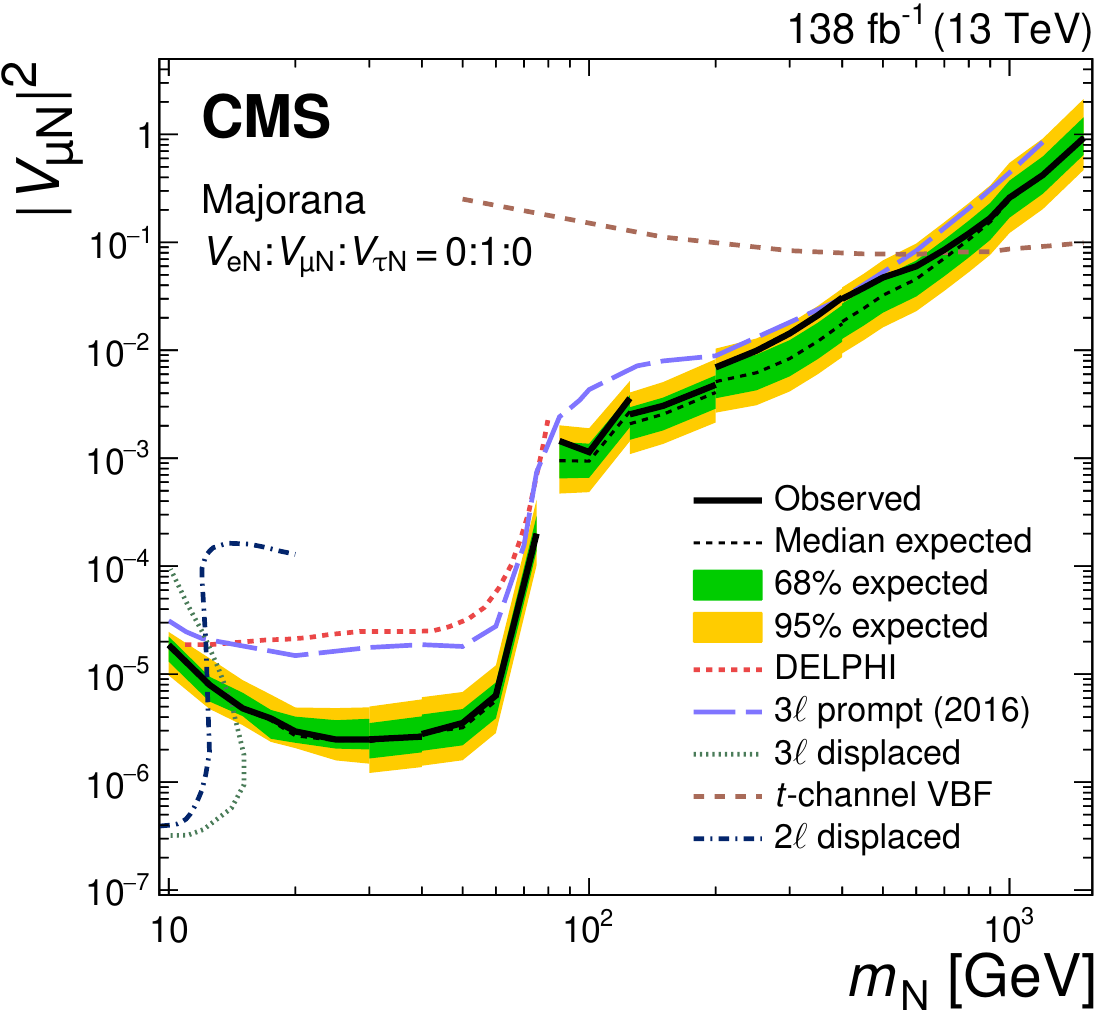}
\hspace{1cm}\includegraphics[width=0.37\textwidth]{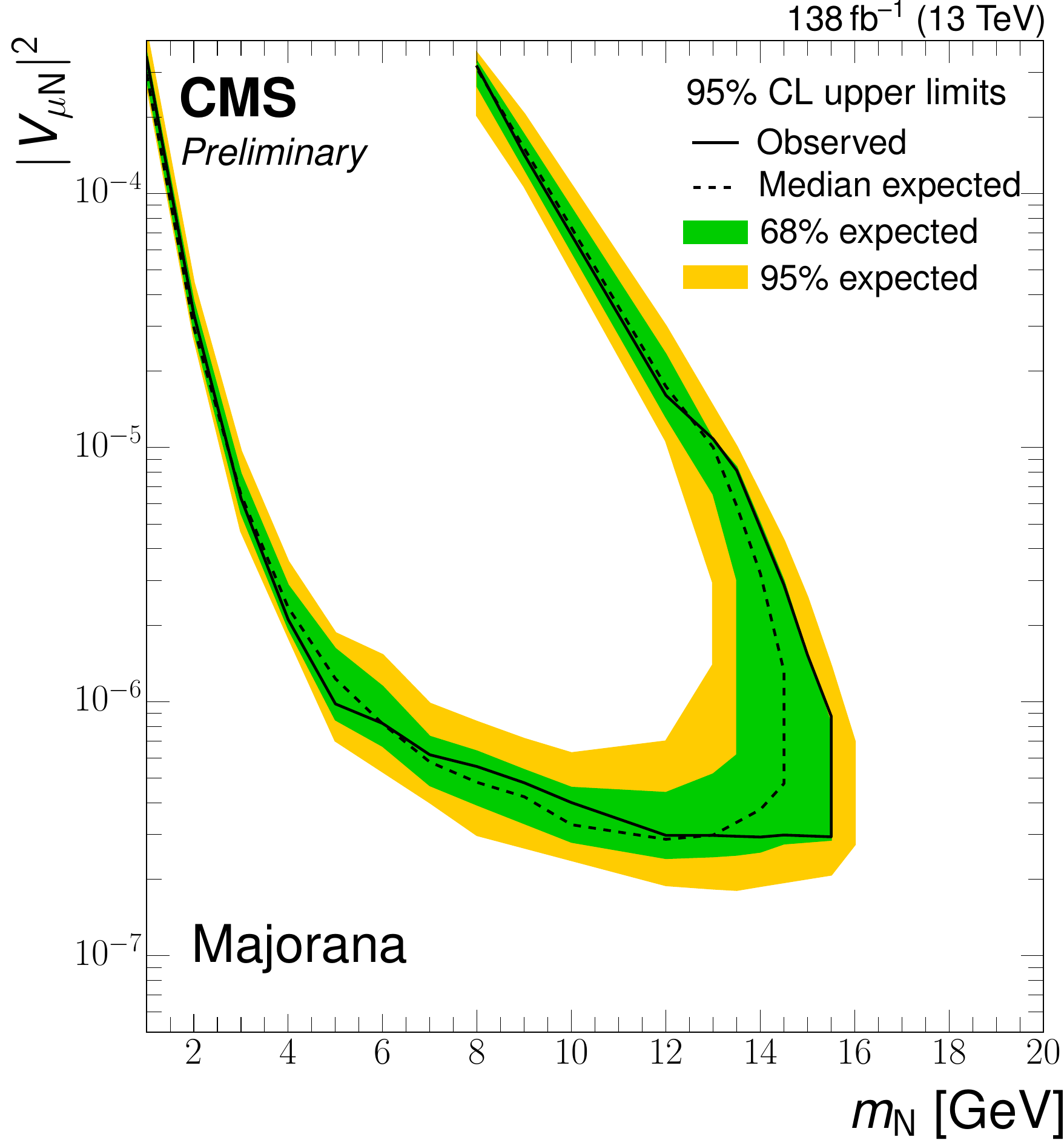}
\caption{The 95\% CL limits on the total mixing amplitude, $|V_{\mu\mathrm{N}}|^2=|V_{\mathrm{N}}|^2$, as a function of the HNL mass $m_\mathrm{N}$, for the short-lived \textit{(left)} and long-lived \textit{(right)} HNL searches in W boson decays. The limits are obtained for the scenario in which the HNL is a Majorana particle that mixes exclusively with the muon sector. \label{figure_WHNL}} 
\end{figure}

Two searches for HNLs in the decays of W~bosons were conducted using Run~2 pp collision data. The first search~\cite{EXO-22-011} assumes that the HNL is short lived and probes final states with three leptons, which can either be electrons, muons, or hadronically decaying taus. The second search~\cite{EXO-21-011}, instead, considers that the HNL is a long-lived particle, such that it travels a measurable distance inside the detector before decaying. In this case, the decay products of the HNL are displaced with respect to the pp interaction point. Final states with a prompt lepton, a displaced lepton, and a displaced jet are analysed with the requirement that leptons are either electrons or muons. In both searches, the observed data are compatible with the SM background prediction. Upper exclusion limits on the total mixing amplitude, $|V_\mathrm{N}|^2$, are presented in Fig.~\ref{figure_WHNL}, for the scenario in which the HNL mixes exclusively with muon neutrinos. Because of the different assumptions on the HNL lifetime, the two searches are sensitive to different mass ranges. The limits are improved up to one order of magnitude compared to previous results. Furthermore, the short-lived HNL analysis presents for the first time limits for the scenario in which the HNL mixes exclusively with tau neutrinos for signal masses greater than the mass of the W boson.

\subsection{Search for long-lived HNLs in B meson decays}\label{section_bhnl}

\begin{figure}[tbh!]
\centering
\includegraphics[width=0.45\textwidth]{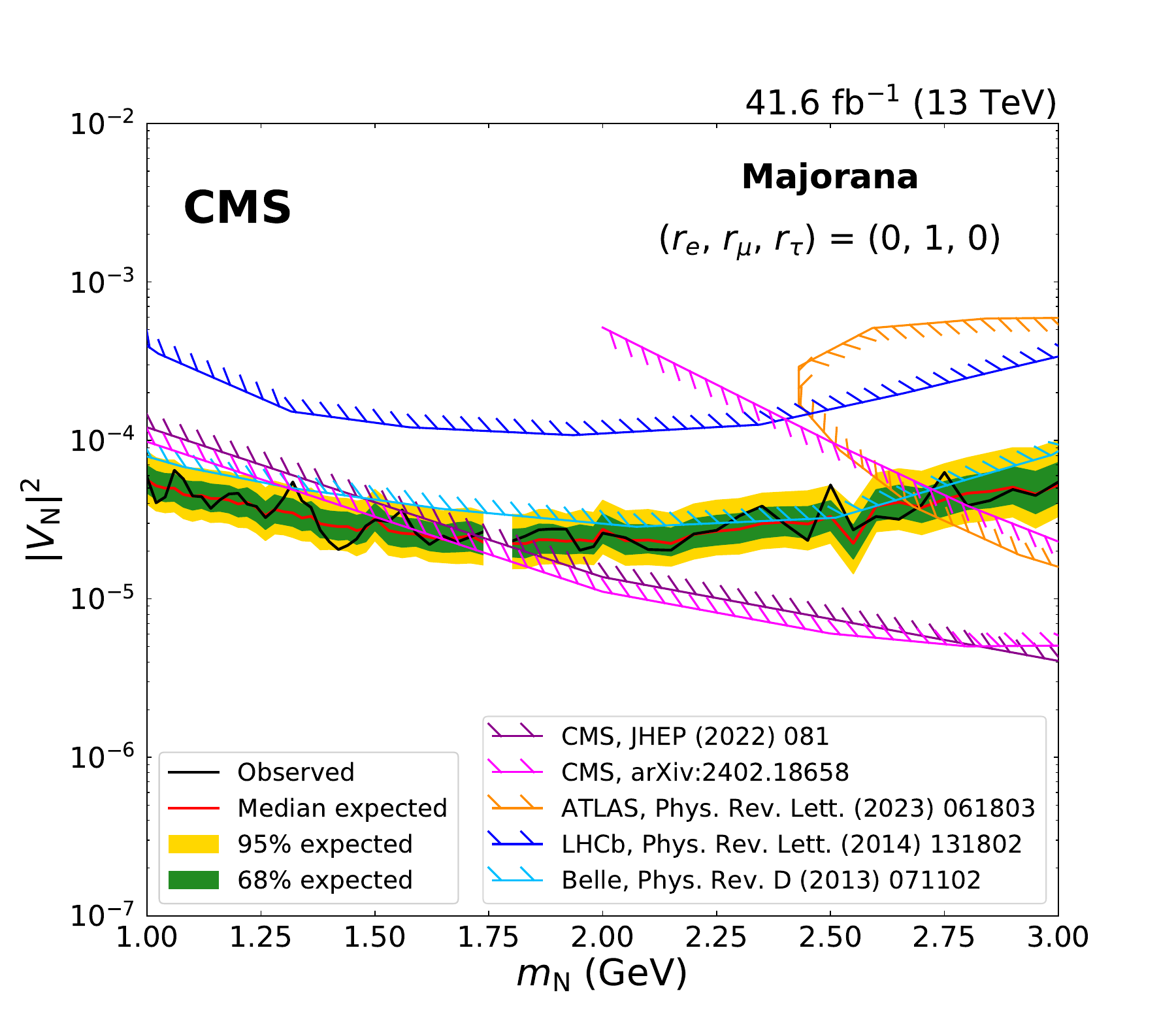}
\includegraphics[width=0.45\textwidth]{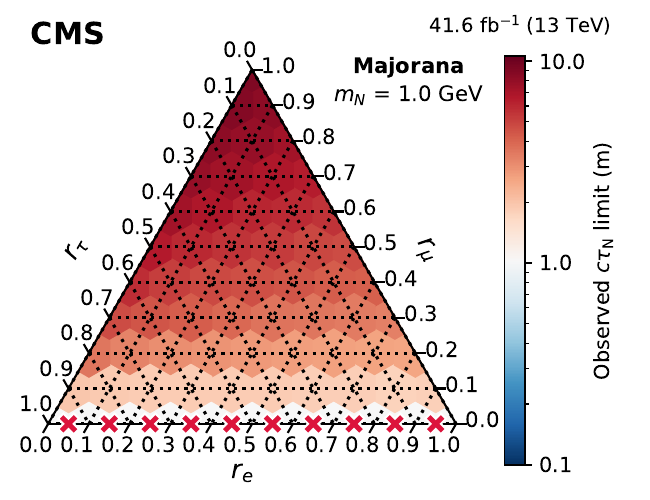}
\caption{\textit{(Left)} The 95\% CL upper limits on the total mixing amplitude, $|V_{\mathrm{N}}|^2$, as a function of the mass $m_\mathrm{N}$ for the HNL search in B meson decays. The limits are obtained for the scenario in which the HNL is a Majorana particle that mixes exclusively with the muon sector. \textit{(Right)} The observed 95\% CL lower limits on the decay length, $c\tau_\mathrm{N}$, as functions of the mixing ratios ($r_\mathrm{e}$, $r_\mu$, $r_\tau$), for a fixed mass $m_\mathrm{N}=1.0$~GeV in the Majorana scenario. \label{figure_BHNL}}
\end{figure}

A search~\cite{EXO-22-019} for long-lived HNLs in the leptonic and semileptonic decays of B~mesons is performed using a special b-hadron-enriched sample, referred to as the B-parking data set~\cite{EXO-23-007} ($\sqrt{s}=13$~TeV, 41.6~fb$^{-1}$), which contains $\mathcal{O}(10^{10})$ b$\overline{\mathrm{b}}$ events. This analysis was notably motivated by the fact that B~mesons are significantly more abundant in pp collisions than W bosons and are therefore, potentially, a more prominent source of HNLs. The process targeted in the search considers B mesons that decay inclusively into a lepton and a long-lived HNL, which, in turn, decays exclusively into a displaced lepton and a displaced pion. The leptons can either be electrons or muons, provided that one of them is a muon that triggers a B-parking line. No significant deviation from the background prediction was observed. Results are interpreted with unprecedented resolution in the HNL mass for various mixing scenarios, specified by different values of the ratios $r_\ell \equiv |V_{\ell\mathrm{N}}|^2/|V_\mathrm{N}|^2$, where $V_{\ell\mathrm{N}}$ is the mixing amplitude to the lepton family  $\ell = (\mathrm{e},\,\mu,\,\tau)$. Upper limits on $|V_\mathrm{N}|^2$ for the scenario in which the HNL mixes exclusively with the muon sector are shown in Fig.~\ref{figure_BHNL} (left). These limits are the most stringent from a collider experiment to date for masses $1.0 < m_\mathrm{N} < 1.7$~GeV. Moreover, lower limits on the HNL decay length, $c\tau_\mathrm{N}$, are provided for 66 mixing scenarios, as shown for $m_\mathrm{N}=1.0$~GeV in Fig.~\ref{figure_BHNL} (right). 

\vspace{-0.03cm}
\section{Model-agnostic search}\label{section_anomaly}

\begin{figure}[tb!]
\centering
\raisebox{-0.5\height}{\includegraphics[width=0.45\textwidth]{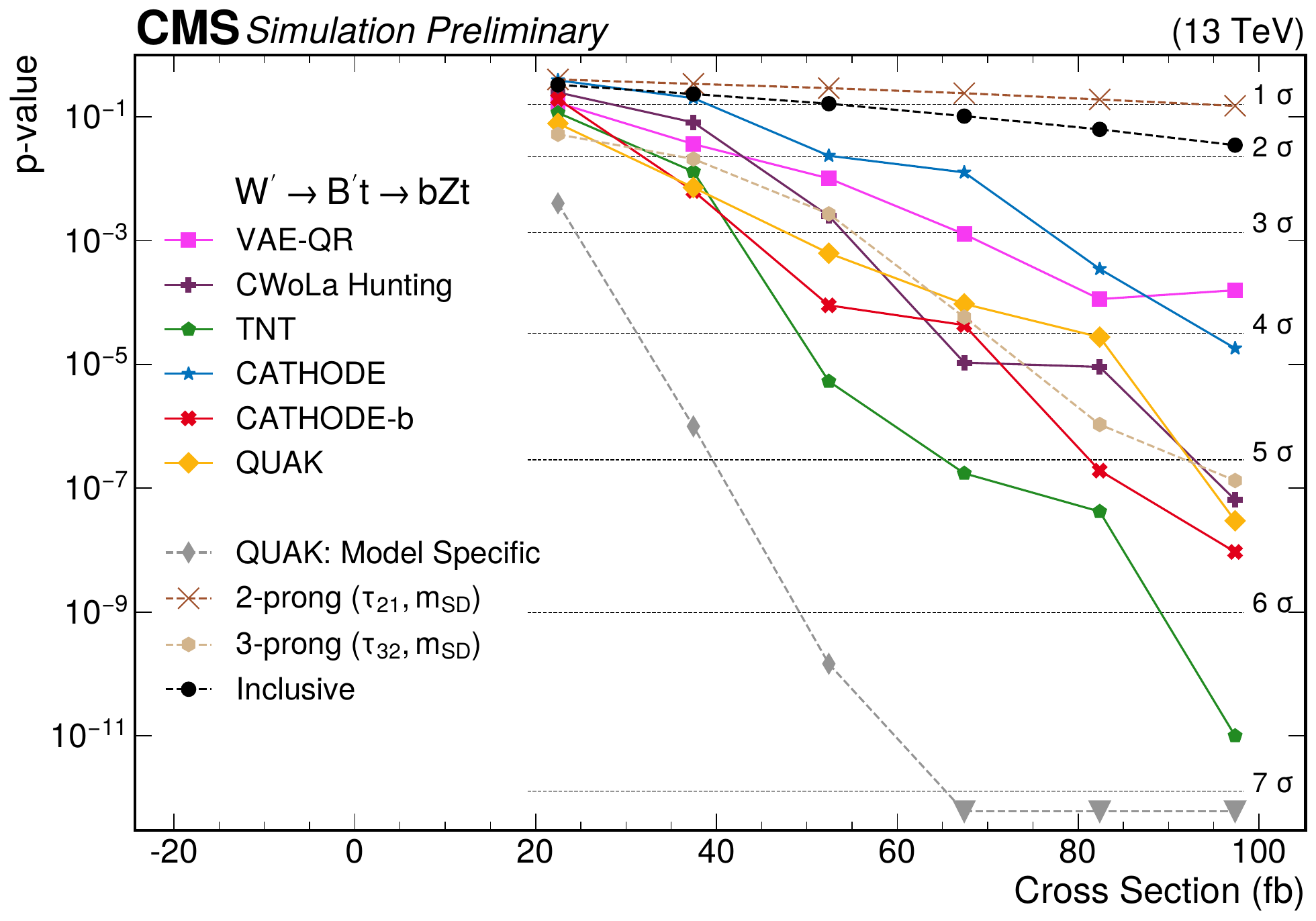}}
\hspace{1cm}
\raisebox{-0.5\height}{\includegraphics[width=0.35\textwidth]{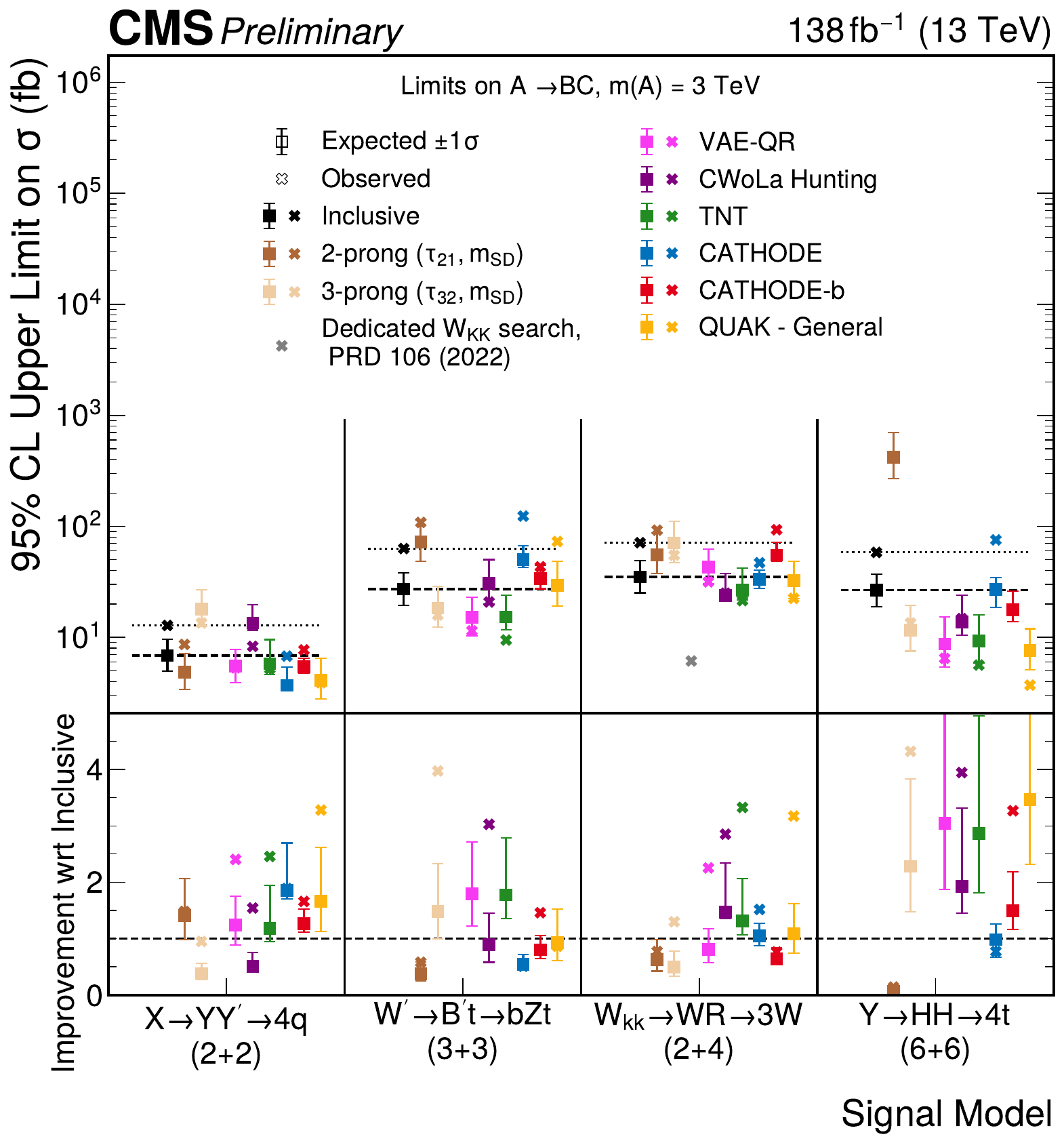}}
\caption{\textit{(Left)} Search p-values as a function of the cross section of an injected 3-pronged signal for the different search methods. The first six labels in the legend indicate the different anomaly detection algorithms and the \textit{inclusive} label refers to results obtained without anomaly detection. \textit{(Right)} The 95\% CL limits on the cross section obtained with the different search methods for four different signal models, shown in the four separate columns. The lower panel shows the ratio of the limits obtained with and without anomaly detection. \label{figure_anomaly}}
\end{figure}

A model-agnostic search~\cite{EXO-22-026} based on anomaly detection was conducted in dijet resonances using Run~2 pp collision data. The generic A$\to$BC process is considered, where A, B, and C are heavy exotic particles, and where B and C decay hadronically into jets. In the case in which the jets originate from a new physics process, it may be assumed that they have an anomalous substructure compared to that of jets originating from an SM process. Because no specific assumption is made on the substructure of the anomalous jets, the sensitivity to unknown new physics signatures is maximised. Five classes of anomaly detection methods based on machine learning were designed and tested in the search. Their discovery power, shown in Fig.~\ref{figure_anomaly} (left), is assessed based on simulation and is promising for all methods. No significant excess of data over the background prediction was observed with any of the anomaly detection methods. The sensitivity of the search is evaluated on benchmark signal models and is found to be enhanced by the anomaly detection, as can be seen from the 95\% CL upper limits shown in Fig.~\ref{figure_anomaly} (right). For most of the inspected signal models, these limits are the first ever presented.

\section{Data parking strategy during Run~3}\label{section_parking}

In view of the success of the B-parking strategy during Run~2, demonstrated in Section~\ref{section_bhnl} by the HNL search in B meson decays, an extended parking program~\cite{EXO-23-007} has been put in place for the Run~3 of data taking. The parking strategy consists in delaying the reconstruction of the recorded events until the computational resources become available. It has proven to be a convenient solution to cope with large event rates resulting from the decrease of the trigger thresholds. During Run~3, B-parking samples will be collected by means of inclusive dimuon and dielectron triggers. Additionally, triggers designed to enhance the sensitivity to vector boson fusion processes, double-Higgs events, and long-lived particles will also use the parking strategy. 

\section{Summary}\label{section_summary}

A great effort is invested at CMS to search for the manifestation of new physics in the data. Recent results from diboson and HNL searches and from a search based on anomaly detection have been presented. No significant excess of data over the SM prediction was observed. Exclusion limits on novel models were established, and already existing limits were improved in many instances. These results owe to the design of advanced data processing and analysis techniques, which are currently being exploited for a successful new physics program during Run~3.


\section*{References}

\begin{thebibliography}{99}
\bibitem{CMS} CMS Collaboration, \href{https://dx.doi.org/10.1088/1748-0221/3/08/S08004}{\color{black}\emph{JINST} \textbf{3}, S08004 (2008)}.
\bibitem{EXO-21-017} CMS Collaboration, \href{https://cds.cern.ch/record/2893032}{\color{black} CMS-PAS-EXO-21-017 (2024)}.
\bibitem{EXO-21-017_hadronic} CMS Collaboration, \href{https://doi.org/10.1016/j.physletb.2022.136888}{\color{black} \emph{Phys. Lett. B} \textbf{826}, 13688 (2022)}.
\bibitem{HIN-21-015} CMS Collaboration, \href{https://cds.cern.ch/record/2895091}{\color{black} CMS-PAS-HIN-21-015 (2024)}.
\bibitem{EXO-22-011} CMS Collaboration, \href{https://arxiv.org/abs/2403.00100}{\color{black} arXiv:2403.00100 (2024)}.
\bibitem{EXO-21-011} CMS Collaboration, \href{https://cds.cern.ch/record/2892670}{\color{black} CMS-PAS-EXO-21-011 (2024)}.
\bibitem{EXO-22-019} CMS Collaboration, \href{https://arxiv.org/abs/2403.04584}{\color{black} arXiv:2403.04584 (2024)}.
\bibitem{EXO-23-007} CMS Collaboration, \href{https://arxiv.org/abs/2403.16134}{\color{black} arXiv:2403.16134 (2024)}. 
\bibitem{EXO-22-026} CMS Collaboration, \href{https://cds.cern.ch/record/2892677}{\color{black} CMS-PAS-EXO-22-026 (2024)}.
\end{thebibliography}
\end{document}